\documentclass[lettersize,journal]{IEEEtran}
\usepackage{amsmath,amsfonts}
\usepackage{algorithmic}
\usepackage{algorithm}
\usepackage{array}
\usepackage[caption=false,font=normalsize,labelfont=sf,textfont=sf]{subfig}
\usepackage{textcomp}
\usepackage{stfloats}
\usepackage{url}
\usepackage{verbatim}
\usepackage{graphicx}
\usepackage{cite}
\usepackage{mathtools}
\usepackage{threeparttable}
\usepackage{makecell}
\usepackage{color}
\usepackage{enumitem}
\begin{document}

\title{Reforming Quantum Microgrid Formation}

\author{Chaofan~Lin,   
        Peng~Zhang, 
		Mikhail A. Bragin,
        and
        Yacov A. Shamash

\thanks{This work relates to Department of Navy award N00014-23-1-2124 issued by the Office of Naval Research. The U.S. Government has a royalty-free license throughout the world in all copyrightable material contained herein.} 
\thanks{C. Lin, P. Zhang and Y. A. Shamash are with the Department of Electrical and Computer Engineering, Stony Brook University, NY 11794, USA (e-mails: chaofan.lin, p.zhang, yacov.shamash@stonybrook.edu).}
 
\thanks{M. A. Bragin is with Southern California Edison, Rosemead, CA 91771, USA (e-mail: mikhail.bragin@sce.com).} 
}

\markboth{
}%
{Shell \MakeLowercase{\textit{et al.}}: A Sample Article Using IEEEtran.cls for IEEE Journals}


\maketitle

\begin{abstract}
This letter introduces a novel compact and lossless quantum microgrid formation (qMGF) approach to achieve efficient operational optimization of the power system and improvement of resilience. This is achieved through lossless reformulation to ensure that the results are equivalent 
to those produced by the classical MGF by exploiting graph-theory-empowered quadratic unconstrained binary optimization (QUBO) that avoids the need for redundant encoding of continuous variables. Additionally, the qMGF approach utilizes a compact formulation that requires significantly fewer qubits compared to other quantum methods thereby enabling a high-accuracy and low-complexity deployment of qMGF on near-term quantum computers. 
Case studies on real quantum processing units (QPUs) empirically demonstrated
that qMGF can achieve the same high accuracy as classic results with a significantly reduced number of qubits. 
\end{abstract}


\section{Introduction}
\IEEEPARstart{M}{icrogrid} formation (MGF) is an effective strategy for boosting distribution system resilience against natural disasters
. Classic MGF is generally formulated as mixed integer linear programming (MILP) with continuous and integer decision variables \cite{My_IJEPES}. However, integer variables result in combinatorial complexity, where the number of possible solutions increases exponentially with the size of the problem, drastically increasing the computation effort \cite{My_IJEPES}.
In recent years, quantum computing has demonstrated 
promise in accelerating the resolution of MGF 
\cite{Nima's_GM, Fei's UC}. 
However, the success of quantum computing methods is contingent on the availability of the quadratic unconstrained binary optimization (QUBO) formulation, which does not account for the presence of continuous variables \cite{Nima's_GM, Fei's UC}. 
To leverage the quantum advantage, 
one common way is to encode the continuous variables with binary ones
\cite{Nima's_GM, Wei's quantum}, which leads to the loss of accuracy as well as to the significant increase of the number of binary variables and quantum-computational requirements.

This letter addressed the above issues at the modeling stages by developing a compact and lossless quantum MGF (qMGF)
that directly formulates the MGF as a QUBO without continuous variables by exploiting the advantages of the graph theory. Rather than resorting to a traditional approach of heuristically determining a redundant mesh of discretization to approximate continuous variables, our novel idea is to establish a new node-to-branch binary decision matrix to explicitly and precisely map the continuous variables in qMGF with existing binary ones.
In doing so, those variables are compactly discretized with a much fewer number of binary variables.\vspace{-5pt}

\section{State-of-the-Practice Quantum Optimization
}

The QUBO solution aims for 
the minimum energy state of the following Ising model \cite{boixo2014evidence}:
\begin{equation}\label{Ising model}
H= -\sum_{j,k}J_{jk}z_jz_k-\sum_jh_jz_j,
\end{equation}
where $H$ is the Hamiltonian function; $z_j$ is the spin variable taking values $\pm1$ ; $J_{jk}$ and $h_j$ are the coefficients.

The problem in (\ref{Ising model}) is also equivalent to finding the ground state over all possible quantum states:
\begin{equation}\label{Final quantum objective}
\min_{|\psi\rangle} \Big\{-\sum_{j,k}J_{jk}\langle \psi|Z_jZ_k|\psi\rangle-\sum_jh_j\langle \psi|Z_j|\psi\rangle\Big\},
\end{equation}
where $|\psi\rangle$ is the quantum state; $Z_jZ_k$ and $Z_j$ are the tensor product of multiple quantum gates, where the indices indicate the positions of each Z gate.

To obtain the ground state (or optimal solution) of \eqref{Final quantum objective}, algorithms 
such as quantum annealing \cite{Nima's_GM} and quantum approximate optimization algorithm (QAOA) \cite{Fei's UC} have been used. For either algorithm, a QUBO formulation is a necessity.



Existing discretization-based methods attempt to approximate such continuous variables in MGF as branch flows and nodal voltages, by a finite number of binary variables \cite{Nima's_GM, Wei's quantum}:
\begin{equation}\label{Encoding}
c=\sum\nolimits_{d=-m_{\rm{F}}}^{m_{\rm{I}}}2^dx_d,
\end{equation}
where $c$ is any continuous variable; $x_d$ is the binary variable for encoding; $m_{\rm{F}}$ and $m_{\rm{I}}$ are the numbers of binary variables to encode the fractional and integer parts, respectively. 

However, the above approximation would inevitably lead to a large number of binary variables, numerical errors, accuracy loss, constraint violations, and infeasibility (See Section IV). To resolve the above issues, an encoding-free compact and lossless QUBO formulation for MGF 
is discussed next.\vspace{-5pt}



\section{A 
Compact and Lossless QUBO Formulation for qMGF}
This section uses quantum notation $|\cdot\rangle$ to denote the binary variables in qMGF\footnote{$z=2x-1 $ should be performed before embedding any binary variable $x$ into actual qubits because $x\in\{0, 1\}$ while $z\in\{-1, 1\}$.}.

\noindent\emph{(1) Microgrid Radial Topology Constraints}

Assuming each formed microgrid (MG) holds a radial topology, the following spanning tree model can be used to partition any structure, 
into MGs with radial topology \cite{My_IJEPES}:
\begin{equation}\label{topology1}
|\alpha_{ij}\rangle=|\beta_{ij}\rangle+|\beta_{ji}\rangle, ij\in\pmb{B},
\end{equation}\vspace{-10pt}
\begin{equation}\label{topology2}
\sum \mathop{}_{ij\in \pmb{B}}|\beta_{ij}\rangle=1, \forall i\in\pmb{N}/\pmb{N}_{\rm{S}},
\end{equation}\vspace{-10pt}
\begin{equation}\label{topology3}
|\beta_{ij}\rangle=0, \forall i\in\pmb{N}_{\rm{S}},
\end{equation}
where $\pmb{B}$ is the set of branches; $\pmb{N}$ is the set of nodes; $\pmb{N}_{\rm{S}}$ is the set of root nodes with power sources; $|\alpha_{ij}\rangle$ is the qubit to decide the status of the branch between nodes $i$ and $j$, where $|\alpha_{ij}\rangle=1$ indicates a closed status or else $|\alpha_{ij}\rangle=0$; $|\beta_{ij}\rangle$ denotes the node relationship, where $|\beta_{ij}\rangle=1$ means node $j$ is the parent node of node $i$ or else $|\beta_{ij}\rangle=0$.

\noindent\emph{(2) Graphical Node-to-Branch-based Network Constraints}

Existing methods formulate the network constraints based on KCL and KVL \cite{ My_IJEPES, Nima's_GM, Fei's UC, Wei's quantum}, which cannot explicitly capture the relationships between the continuous power flows/nodal voltages and discrete load/branch statuses. To explicitize the relationships, instead of KCL and KVL, we define a new node-to-branch (N2B) decision matrix from a graphical perspective:
\begin{equation}\label{N2B}
N2B\coloneqq |\pi_{i\rightarrow jk}\rangle, i\in \pmb{N}, jk\in \pmb{B},
\end{equation}
which equals 1 if the path between node $i$ and any root node passes through branch $jk$ and equals 0 if not.
For a radial MG graph, the following constraints should be satisfied:
\begin{equation}\label{N2B}
\left\{
\begin{aligned}
|\pi_{i\rightarrow jk}\rangle\geq |\pi_{h\rightarrow jk}\rangle|\alpha_{ih}\rangle\\
|\pi_{h\rightarrow jk}\rangle\geq |\pi_{i\rightarrow jk}\rangle|\alpha_{ih}\rangle
\end{aligned}
, ih\in \pmb{B}/jk, jk\in \pmb{B},\right.
\end{equation}
i.e., if branch $ih$ is closed, then nodes $i$ and $h$ share the same pass-through branch $jk$ or not at the same time; if open, their passing statuses through branch $jk$ have no relationship. The quadratic term $|\pi\rangle|\alpha\rangle$ can be linearized by the method in \cite{Nima's_GM}.

For nodes and their directly connected branches, we have:
\begin{equation}\label{N2B_same1}
\left\{
\begin{aligned}
|\pi_{i\rightarrow ih}\rangle=|\beta_{ih}\rangle\\
|\pi_{h\rightarrow ih}\rangle=|\beta_{hi}\rangle
\end{aligned}
, ih\in \pmb{B},\right.
\end{equation}
i.e., only the child node passes through its connected branch.

Moreover, the root nodes should pass through no branches:
\begin{equation}\label{N2B_root}
|\pi_{i\rightarrow jk}\rangle=0, i\in \pmb{N}_{\rm{S}}, jk\in \pmb{B}.
\end{equation}

With the N2B matrix, the active and reactive power flow and voltage drop at each branch can be explicitly expressed by:\vspace{-5pt}
\begin{equation}\label{Branch P}
P_{jk}=\sum \mathop{}_{i\in\pmb{N}}|\lambda_{i}\rangle|\pi_{i\rightarrow jk}\rangle P^{\rm{L}}_{i}, jk\in \pmb{B},
\end{equation}
\begin{equation}\label{Branch Q}
Q_{jk}=\sum \mathop{}_{i\in\pmb{N}}|\lambda_{i}\rangle|\pi_{i\rightarrow jk}\rangle Q^{\rm{L}}_{i}, jk\in \pmb{B},
\end{equation}
\begin{equation}\label{Branch U}
\begin{aligned}
&\Delta U_{jk}=U_{k}-U_{j}=(R_{jk}P_{jk}+X_{jk}Q_{jk})/U_0=\\
&\sum \mathop{}_{i\in\pmb{N}}|\lambda_{i}\rangle|\pi_{i\rightarrow jk}\rangle(R_{jk}P^{\rm{L}}_{i}+X_{jk}Q^{\rm{L}}_{i})/U_0, jk\in \pmb{B},
\end{aligned}
\end{equation}
where $|\lambda_{i}\rangle$ is the qubit to decide whether to restore the load at node $i$ or not; $P_{jk}$ and $Q_{jk}$ are the active and reactive power flows from node $j$ to $k$; $U_{j}$, $U_{k}$, $\Delta U_{jk}$, and $U_0$ are the voltages at nodes $j$ and $k$, the voltage drop from node $k$ to $j$, and the nominal voltage; $P^{\rm{L}}_{i}$ and $Q^{\rm{L}}_{i}$ are the active and reactive powers of the load at node $i$; $R_{jk}$ and $X_{jk}$ are the resistance and reactance of branch $jk$.

\noindent\emph{(3) Security Constraints}

Based on (\ref{Branch P}) to (\ref{Branch U}), the security constraints of all sources, branches, and nodes in qMGF can be formulated as:
\begin{equation}\label{Source P}
P_{j}^{\rm{min}}\leq|\lambda_{j}\rangle P^{\rm{L}}_{j}+\sum_{k\in j}\sum_{i\in\pmb{N}}|\lambda_{i}\rangle|\pi_{i\rightarrow jk}\rangle P^{\rm{L}}_{i}\leq P_{j}^{\rm{max}}, j\in \pmb{N}_{\rm{S}},
\end{equation}
\begin{equation}\label{Source Q}
Q_{j}^{\rm{min}}\leq|\lambda_{j}\rangle Q^{\rm{L}}_{j}+\sum_{k\in j}\sum_{i\in\pmb{N}}|\lambda_{i}\rangle|\pi_{i\rightarrow jk}\rangle Q^{\rm{L}}_{i}\leq Q_{j}^{\rm{max}}, j\in \pmb{N}_{\rm{S}},
\end{equation}
\begin{equation}\label{Branch P limit}
\sum \mathop{}_{i\in\pmb{N}}|\lambda_{i}\rangle|\pi_{i\rightarrow jk}\rangle P^{\rm{L}}_{i}\leq |\alpha_{jk}\rangle P_{jk}^{\rm{max}}, jk\in \pmb{B},
\end{equation}\vspace{-12pt}
\begin{equation}\label{Branch Q limit}
\sum \mathop{}_{i\in\pmb{N}}|\lambda_{i}\rangle|\pi_{i\rightarrow jk}\rangle Q^{\rm{L}}_{i}\leq |\alpha_{jk}\rangle Q_{jk}^{\rm{max}}, jk\in \pmb{B},
\end{equation} \vspace{-10pt}
\begin{equation}\label{Node U limit}
\begin{aligned}
\sum \mathop{}_{jk\in \pmb{B}}|\pi_{h\rightarrow jk}\rangle \sum \mathop{}_{i\in\pmb{N}}|\lambda_{i}\rangle|\pi_{i\rightarrow jk}\rangle (R_{jk}P^{\rm{L}}_{i}+X_{jk}Q^{\rm{L}}_{i})/U_0\\
\leq \Delta U_{h}^{\rm{max}}+(1-\lambda_{h})U_\delta,h\in\pmb{N},
\end{aligned}
\end{equation}
where $P_{j}^{\rm{min}}$ and $P_{j}^{\rm{max}}$ ($Q_{j}^{\rm{min}}$ and $Q_{j}^{\rm{max}}$) are the minimum and maximum active (reactive) power outputs of the source at node $j$; $k\in j$ denotes node $k$ is connected to node $j$; $P_{jk}^{\rm{max}}$ and $Q_{jk}^{\rm{max}}$ are the maximum active and reactive power flows of branch $jk$; $\Delta U_{h}^{\rm{max}}$ is the maximum voltage drop for node $h$; $U_\delta$ is a small voltage value, which relaxes the upper boundary when the load at node $h$ is not restored ($|\lambda_{h}\rangle=0$).

\emph{(4) Objective of qMGF}

The objective of qMGF, same as classic MGF, is to maximize the restored load amount considering priorities \cite{My_IJEPES, Nima's_GM, Fei's UC}:
\begin{equation}\label{objective}
obj=\max_{|\lambda_{i}\rangle, |\alpha_{ij}\rangle, |\beta_{ij}\rangle, |\pi_{i\rightarrow jk}\rangle}{\sum\mathop{}_{i\in\pmb{N}}|\lambda_{i}\rangle w_i P^{\rm{L}}_{i}},
\end{equation}
where $w_i$ is the weight of the load at node $i$.

It is worth mentioning that, although 
qMGF introduces a new N2B decision matrix, it usually has a sub-quadratic to linear complexity with increased system scale (number of nodes). This is because in most cases the set of possible power supply paths for a certain node does not always cover all branches of the system, and thus the N2B matrix is not full or even sparse.
To determine which elements are not decision variables, one can perform the simple path search in the graph theory for each node and all sources and find out those branches that the node would impossibly pass through.  

The above formulation is a binary optimization that needs to be converted to a QUBO formulation before a QC can solve it. The conversion includes 1) Float coefficients in (\ref{Source P}) to (\ref{Node U limit}) to integer ones by multiplication by $10^n$ on both sides of the constraints;
2) Inequality constraints to equality ones by adding slack variables; and 3) Equality constraints to the objective by adding tuned penalty coefficients. The details of this conversion can be found in \cite{Nima's_GM}. Overall algirithm is below:
\vspace{-12pt}
\begin{algorithm}[H]
\caption{qMGF Algorithm.}\label{algorithm}
\begin{algorithmic}
\STATE \textbf{Input:} $P^{\rm{L}}_{i}$, $Q^{\rm{L}}_{i}$, $w_i$, $\Delta U_{i}^{\rm{max}}$, $R_{ij}$, $X_{ij}$, $P_{ij}^{\rm{max}}$, $Q_{ij}^{\rm{max}}$, $P_{i}^{\rm{min}}$, $P_{i}^{\rm{max}}$, $Q_{i}^{\rm{min}}$, $Q_{i}^{\rm{max}}$, $\pmb{B}$, $\pmb{N}$, $\pmb{N}_{\rm{S}}$
\STATE \textbf{Do} qMGF formulation with (\ref{topology1})-(\ref{topology3}), (\ref{N2B})-(\ref{N2B_root}), (\ref{Source P})-(\ref{objective})
\STATE \ \ \ \ convert the formulation to QUBO
\STATE \ \ \ \ solve the QUBO by D-Wave QPU and obtain the values of $|\lambda_{i}\rangle$, $|\alpha_{ij}\rangle$, and $|\pi_{i\rightarrow jk}\rangle$ corresponding to the ground state
\STATE \ \ \ \ calculate branch flows and nodal voltages by (\ref{Branch P}) to (\ref{Branch U})
\STATE \textbf{Results:} $\lambda_{i}$, $U_i$, $\alpha_{ij}$, $P_{ij}$, $Q_{ij}$
\end{algorithmic}
\label{alg1}
\end{algorithm}
\vspace{-10pt}
\section{Case Study}

\subsection{Accuracy Advantage}

The accuracy of qMGF is validated on a modified IEEE 37 node test feeder \cite{My_IJEPES} using Gurobi 11.0.
Table \ref{comparison} compares the results of qMGF, classic MGF with an MILP formulation \cite{My_IJEPES} (cMGF) and encoding discretization-based MGF with a QUBO formulation \cite{Nima's_GM, Wei's quantum} (dMGF). 

\begin{table}[!ht]
\vspace{-10pt}
\caption{Comparison of qMGF with Encoding Discretization-based MGF and Classic MGF\label{comparison}}
\centering
\begin{threeparttable}
\begin{tabular}{p{3cm}<{\centering}|p{0.7cm}<{\centering}p{0.7cm}<{\centering} p{0.7cm}<{\centering} p{0.7cm}<{\centering} p{0.7cm}<{\centering}}
\hline
Item & \textbf{qMGF} & dMGF1\tnote{1} & dMGF2 & dMGF3 & cMGF\\
\hline
No. of con. vars. & \textbf{0} & 0 & 0 & 0 & 122  \\
No. of bin. vars./qubits & \textbf{4134} & 8409 & 10901 & 11257 & 141 \\
Objective value (e6) & \textbf{2.3095} & 2.4425 & 2.4181 & 2.3095 & 2.3095 \\
Load served ratio\tnote{2} (\%) & \textbf{52.0414} & 55.7005 & 54.4326 & 52.0414 & 52.0414\\
Constraint violation sum\tnote{3} & \textbf{0} & 14.5397 & 9.2309 & 2.1152 & /\\
Ave. con. vars. error (\%) & \textbf{0} & 7.5183 & 3.1903 & 0.0008 & /\\
\hline
\end{tabular}
\begin{tablenotes}
\item[1] dMGF1, dMGF2, and dMGF3 denote the dMGFs in different encoding granularities (total numbers of qubits {(8268, 10760, and 11116)} for encoding continuous variables).
\item[2] This means the active power ratio of the served load to the total load \cite{My_IJEPES}.
\item[3] The constraint violation sum is calculated by substituting all variable values corresponding to the optimal solution into the original MILP formulation (cMGF) and summing up all constraint violations.
\end{tablenotes}
\end{threeparttable}
\end{table}

It shows that: 
\begin{itemize}[leftmargin=*]
    \item qMGF needs 
    only 35\%-50\% of qubits that dMGF needs to achieve the same accuracy. 
    \item qMGF obtains the same optimal solution as cMGF does without constraint violations or variable errors.
    \item dMGF inevitably introduces numerical errors that could lead to constraint violations and variable errors. These errors can be reduced 
    at an expensive, oftentimes prohibitive, price of increasing the number of qubits.
\end{itemize}

\begin{figure}
\centering
\includegraphics[width=2.8in]{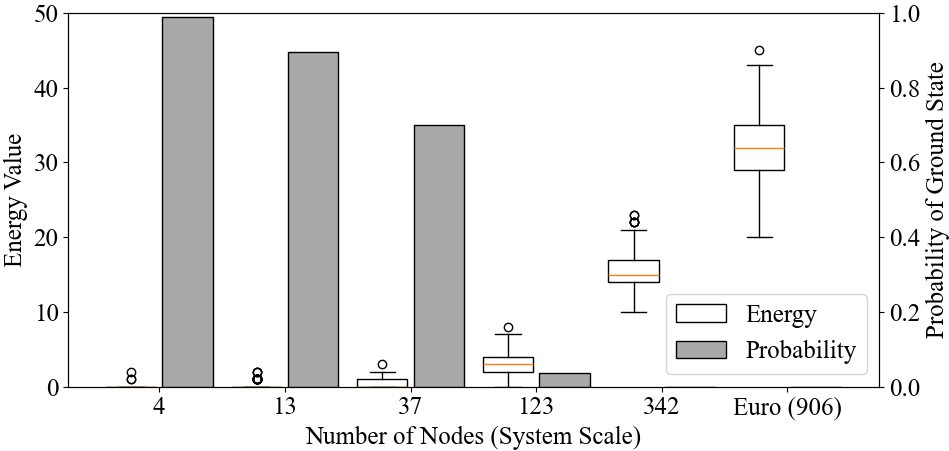}
\caption{The distributions of energy values of 300 samples and the probabilities of the ground state for topology optimization in different system scales.}
\label{Fea_qpu_top_fig}
\end{figure}

\begin{figure}
\centering
\includegraphics[width=2.9in]{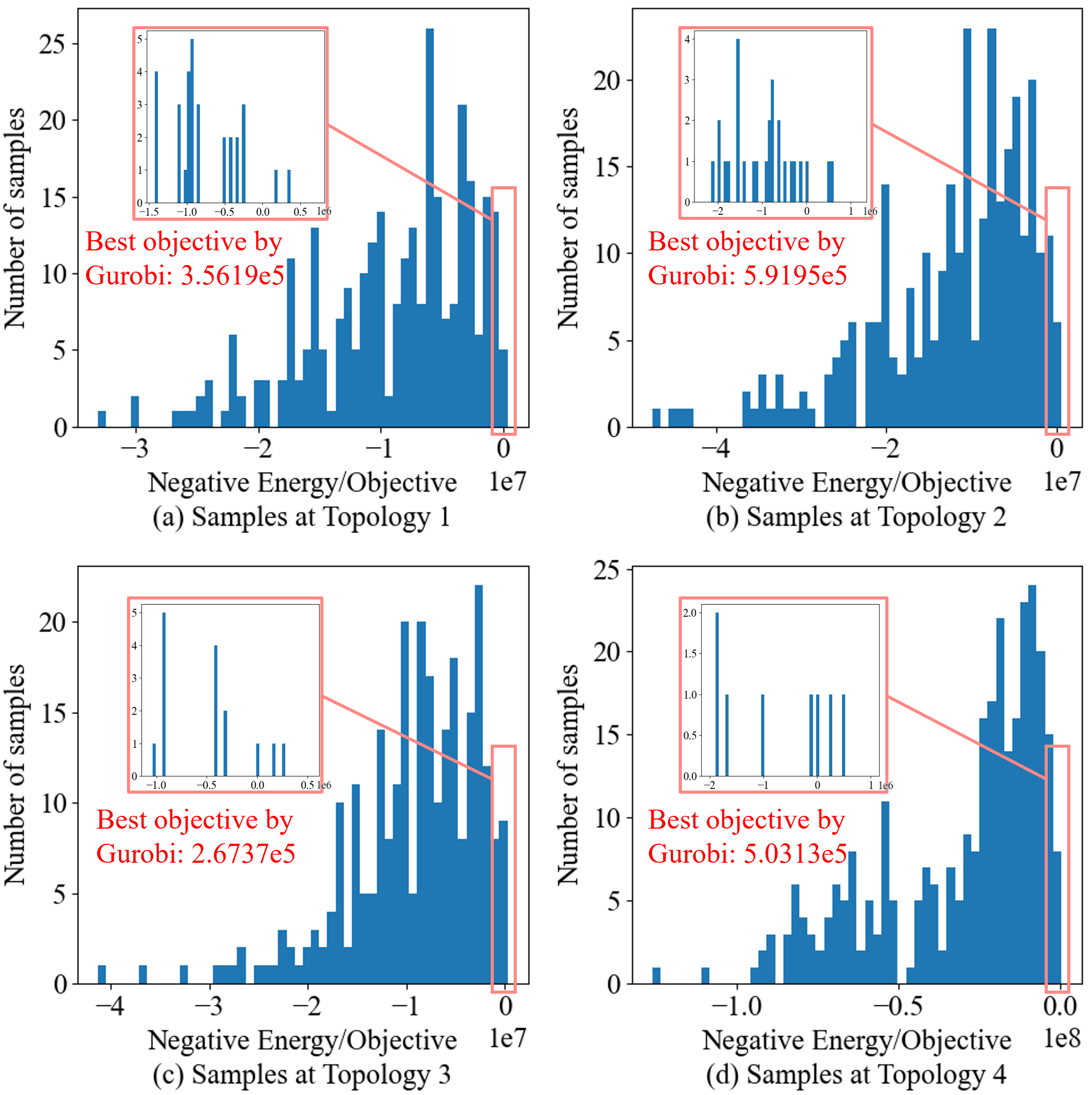}
\caption{The distributions of energy values of 300 samples for restoration optimization at different topologies of the 4 node system.}
\label{Fea_qpu_fig}
\end{figure}

 Therefore, qMGF outperforms dMGF with higher accuracy and reduced number of qubits required.
\vspace{-5pt}
\subsection{Performance on Real QPU}

qMGF is deployed and evaluated on a D-Wave's QPU solver \emph{Advantage\_system6.4} with 5760 qubits. Six IEEE PES test feeders at different scales (4, 13, 37, 123, 342, and 906 node systems) are selected for the evaluation. The key results are:
\begin{itemize}[leftmargin=*]
    \item Due to the scale and noise issues in real QPU, it failed to sample out the ground state or optimal solution in a limited time or sample size (e.g., 1e6 samples), even if for the smallest 4-node system that only needs 114 qubits.
    \item For topology optimization, Fig. \ref{Fea_qpu_top_fig} shows that, as system scale increases, the energy values of samples tend to deviate from the lowest zero and the probability of successfully sampling out the ground state decreases correspondingly.
    \item For restoration optimization, as shown in Fig. \ref{Fea_qpu_fig}, the ground states at different topologies are all successfully sampled out in 300 samples, indicating the potential feasibility of qMGF on real QPU.
\end{itemize}

It is noted that some existing commercial solvers (e.g., CQM for D-Wave) handle continuous variables or MILP through proprietary quantum-classic 
hybrid schemes where the task assignments are largely invisible to users. This letter seeks to provide a reformed general qMGF method that is poised to be run on genuine QCs other than hybrid solvers; thus our approach is platform-independent, readily applicable on 
any 
available QC platforms. 

\section{Conclusion}

This letter presents a compact and lossless quantum microgrid formation (qMGF) method to accurately and efficiently solve the MGF problem on real QCs. qMGF achieves the same accuracy as that of classic MGF, whereas its new problem formulation requires fewer qubits and leads to lower computational complexity than the vanilla quantum methods. Thus it has promising potential to be deployed on the noisy-intermediate-scale quantum computers. A future direction is to further accelerate qMGF for real-scale distribution systems with inverter-based resources.

\vfill

\end{document}